%% file: main.tex
\documentclass[aps,physrev,twocolumn,superscriptaddress]{revtex4-2}
\usepackage{orcidlink}
\usepackage{amsmath}
\newcommand{\multilinecell}[2][c]{%
  \begin{tabular}[#1]{@{}c@{}}#2\end{tabular}}
\newcommand{\multilinecellleft}[2][c]{%
  \begin{tabular}[#1]{@{}l@{}}#2\end{tabular}}

\begin{document}

\include{Macros/parameters}

%\preprint{}

%\title{Impact of arm orientation choice}
\title{So Happy Together: Einstein Telescope and Cosmic Explorer can search for both compact binary and stochastic background signals}

\author{Michael Ebersold\,\orcidlink{0000-0003-4631-1771}}
\email[]{michael.ebersold@physik.uzh.ch}
\affiliation{Physik-Institut, University of Zurich, Winterthurerstrasse 190, 8057 Zurich, Switzerland}

\author{Tania Regimbau}
\affiliation{Laboratoire d'Annecy de Physique des Particules, CNRS, 9 Chemin de Bellevue, 74941 Annecy, France}

\author{Nelson Christensen\, \orcidlink{0000-0002-6870-4202}}
\affiliation{Universit\'e C$\hat{o}$te d’Azur, Observatoire de la C$\hat{o}$te d’Azur, CNRS, Artemis, 06304 Nice, France}

\date{\today}

\begin{abstract}
A proposal for the Einstein Telescope (ET) gravitational-wave detector is to construct two L-shaped detectors in Europe, separated by $\sim 1000$ km. The relative orientation of these detectors is critical for their ability to observe a stochastic gravitational-wave background. Aligned detectors, corresponding to a relative orientation of $0^\circ$, offer the best stochastic background sensitivity, whereas a relative orientation of $45^\circ$ improves the measurement of the two gravitational-wave polarizations, and yields a slightly better detection probability and estimate of the luminosity distance for compact binary coalescence; the sky-localization is similar for the two cases. We show that intermediate orientations provide a good compromise between stochastic background and compact binary coalescence searches. The addition of a single Cosmic Explorer (CE) detector in the US considerably improves compact-binary-coalescence observations, in particular localization through triangulation. With CE, the network can measure both polarizations, regardless of the orientation of the two ET Ls. For stochastic background searches, however, the two ET Ls still require a good relative alignment, as the large distance to CE causes its contribution to fall off quickly with frequency. The planning for the global network of next generation gravitational-wave detectors must carefully consider detector orientation to ensure that both compact binary coalescence and stochastic gravitational-wave background searches are effective. Compromise orientations are possible.
\end{abstract}

%\keywords{}

\maketitle

\section{\label{sec:Introduction} Introduction}
The LIGO--Virgo--KAGRA (LVK) Collaboration has achieved much success to date with the announcement of $\sim 390$ gravitational-wave events~\cite{LIGOScientific:2026wfs}. All of these events have been from compact binary coalescences (CBCs). There have been two signals from binary neutron star (BNS) mergers~\cite{LIGOScientific:2017vwq,LIGOScientific:2020aai}, several from neutron star-black hole mergers~\cite{LIGOScientific:2021qlt,LIGOScientific:2024elc,LIGOScientific:2020zkf}, with the rest from binary black hole (BBH) mergers. 

These observations have been made by the two Advanced LIGO detectors~\cite{LIGOScientific:2014pky}, and the Advanced Virgo detector~\cite{VIRGO:2014yos}. The two LIGO interferometric detectors in the US (LIGO Livingston, Louisiana; LIGO Hanford, Washington) are separated by 3000 km, and have nearly identical orientations, with each detector rotated by the same angle relative to the great circle connecting the two sites. The Virgo detector is 7912 km from LIGO Livingston, or 77$^\circ$ of arc length about the Earth surface. As such, Virgo has a different response to the gravitational-wave polarizations than the LIGO detectors. It is this network that has achieved stunning success in detecting CBC-produced gravitational-wave signals. A three detector network is able to estimate the location of a source on the sky, as was the case for the observation of the BNS produced GW170817~\cite{LIGOScientific:2017vwq}. The improved sky localization provided by the three-detector observation of GW170817 then led to the detection of the associated kilonova~\cite{LIGOScientific:2017ync}. 

The LVK searches for other types of gravitational-wave signals. Un-modeled transient signals, or "bursts", could be produced, for example, from supernovae~\cite{LIGOScientific:2019ryq}, pulsar glitches~\cite{LIGOScientific:2025dup}, or cosmic strings~\cite{LIGOScientific:2021nrg}.
Rapidly rotating neutron stars, if they have a non-axisymmetric triaxial deformation, will emit gravitational waves that could be detected by the LVK network~\cite{LIGOScientific:2026qsb}.

A stochastic gravitational-wave background (SGWB) is a random, incoherent superposition of unresolvable gravitational-wave signals produced by the cumulative emission of  astrophysical sources and fundamental cosmological processes integrated across cosmic time~\cite{Christensen:2018iqi}. Direct detection of a cosmological stochastic gravitational-wave background would open an unattenuated observational window into the ultra-early Universe, probing physical processes at energy scales far exceeding those attainable by terrestrial particle accelerators~\cite{Caprini:2018mtu,Christensen:2018iqi}. Such a discovery would yield transformative insights into high-energy phenomena, such as cosmic inflation, first-order cosmological phase transitions, and networks of cosmic strings, providing decisive observational benchmarks for fundamental physics beyond the Standard Model~\cite{Maggiore:1999vm}. LIGO and Virgo have already set impressive constraints on various theories through their limits on the strength of the SGWB~\cite{LIGOScientific:2025kry}. 

The dimensionless energy density spectrum $\Omega_{\text{GW}}(f)$ quantifies the fraction of the Universe's critical energy density $\rho_c$ contained within the SGWB per unit logarithmic frequency, defined as $\Omega_{\text{GW}}(f) = \frac{1}{\rho_c} \frac{\mathrm{d}\rho_{\text{GW}}}{\mathrm{d}\ln f}$. The first attempt by LIGO (in 2004) to detect an SGWB led to an upper limit of 
$\Omega_\text{GW} < 51$~\cite{LIGOScientific:2003jxj}. Twenty years later, the upper limit has been reduced to $\Omega_\text{GW} < 2.8 \times 10^{-9}$~\cite{LIGOScientific:2026mjf}. This is an impressive improvement of a factor $1.8 \times 10^{10}$. Such an upper limit has already been used to constrain a number of cosmological and high energy physics theories~\cite{LIGOScientific:2025kry}. 

An astrophysically produced SGWB arises from the incoherent, unresolvable superposition of countless CBCs, primarily BBH and BNS mergers, occurring throughout cosmic history~\cite{LIGOScientific:2016fpe,LIGOScientific:2017zlf,Christensen:2018iqi}. Observing this background with the LVK detector network would provide a powerful probe of compact binary populations beyond the horizon of individual, high signal-to-noise detections~\cite{LIGOScientific:2016fpe,Callister:2020arv}. Crucially, constraining its energy density spectrum $\Omega_{\text{GW}}(f)$ would yield vital insights into the cosmic star formation rate, merger rate density evolution, and average stellar mass distribution across cosmic time~\cite{Regimbau:2011rp,Regimbau:2022mdu,Ebersold:2025izh}. An astrophysically produced SGWB could be detected by the LVK network during the upcoming O5 observing run~\cite{LIGOScientific:2026mjf}.

The next generation of ground based gravitational-wave detectors, such as the Einstein Telescope (ET)~\cite{Punturo_2010,ET:2025xjr} and Cosmic Explorer (CE)~\cite{Reitze:2019iox}, is currently being planned. The original ET design consisted of an equilateral triangle with 10 km arms and six interferometers; comprising three low-frequency and three high-frequency interferometers. There is now consideration for ET to consist of two L-shaped detectors, with arm lengths of 15 km~\cite{Branchesi:2023}. Each L would consist of a low frequency interferometer and a high frequency interferometer. The two Ls would be in Europe, and displaced from each other by $\sim 1000$ km. CE would consist of a single L-shaped interferometer, with 40 km length arms; it would be located in the US. There is the possibility that there could be a second CE interferometer, with 20 km length arms. Unlike detector sensitivity, which is largely fixed by the instrument design, the relative orientations of these detectors within a network remain an important degree of freedom that influences both CBC and SGWB science.

In this work, we specifically discuss two ET Ls in Europe, separated by 1250 km, and a single CE in the US. We study the orientation of the two ET Ls, and how it influences the possible detection of a SGWB, or the detection and localization of CBC events. These objectives place competing requirements on the detector orientations, motivating a quantitative study of the trade-offs involved. The key quantity is the relative orientations of the two L's. To define it, one first introduces the great circle connecting the detector vertices~\cite{PhysRevD.46.5250,PhysRevD.55.448}. A relative orientation 0$^\circ$ corresponds to the two detectors having identical orientations with respect to this great circle. In this case, the responses of the detectors overlap maximally, providing the optimal configuration for searches for an isotropic SGWB. The sensitivity to a SGWB is further aided by having the distance between the two Ls as small as possible. In this way the overlap reduction function falls off slower with frequency~\cite{PhysRevD.46.5250,Flanagan:1993ix,PhysRevD.55.448}. At a relative orientation of 45$^\circ$, however, the overlap reduction function is zero, and there is no sensitivity to a SGWB~\cite{Christensen:2026yyl}.

For CBC observations, there is a preference for relative orientations close to 45$^\circ$ between the two detectors~\cite{Branchesi:2023}. In this configuration, the network is more sensitive to both gravitational-wave polarizations, which improves its ability to reduce the degeneracy between the luminosity distance and the binary inclination~\cite{Cutler:1994ys}. Nevertheless, even for a relative orientation of 0$^\circ$, sky localization is comparable to that of the 45$^\circ$ orientation, as we demonstrate below.
With a separation of only 1250 km, two ET-L detectors in Europe have limited ability to localize CBC sources through triangulation. This baseline is approximately 2.4 times shorter than that between the two LIGO observatories. As a result, the sky-localization performance is similar for relative orientations of 0$^\circ$ and 45$^\circ$. As we show below, the addition of a CE detector in the US allows for good triangulation and a much better determination of the source localization on the sky.

In this work, we apply the network-optimization framework developed in Ref.~\cite{Ebersold:2024} to a network consisting of two ET Ls in Europe and a CE detector in the US. We identify the orientations that maximize CBC and SGWB performance and, in particular, determine configurations that provide a balance between these two science goals. We then investigate the impact of these orientations on CBC and SGWB observations.
To assess the role of the CE detector, we compare the performance of the full three-detector network with that of the ET-only network. While the addition of CE substantially improves CBC detection and source localization, the sensitivity to an isotropic SGWB remains strongly dependent on the alignment of the two ET Ls, since a third detector located far from Europe will not provide much improvement to the stochastic sensitivity. As such, it is critically important that the alignment of the two ET Ls in Europe is chosen carefully. 

We note that the Einstein Telescope Organization has recently released their Geometry Comparison Report~\cite{CERN:EDMS-3529996}. They compared the ET triangle (10 km arm lengths) with two Ls (15 km arm lengths). The 2 L network had an orientation of 42.5$^\circ$; another orientation considered was 2.5$^\circ$. An intermediate orientation of 30$^\circ$ was studied for CBC detection. For the 2 L network the separation was 1200 km. The report presented SGWB sensitivities, CBC detection efficiencies, and CBC source position estimates. The inclusion of CE was studied for CBC signals~\cite{CERN:EDMS-3529996_Annex_3}.
Another recent study has investigated the mutual detection of CBCs and a SGWB with ET~\cite{Bruno:2026rwf}. This study considers the ET triangle along with an ET L in Europe, and also an additional CE L in the US. The results of that study are consistent with what we present here.

The organization of this publication is as follows: in Sec.~\ref{sec:Networks} we describe the strategy to choose the orientation angles and define the detector-network configurations considered in this study. Then we investigate the performance of these networks regarding CBC searches in Sec.~\ref{sec:ImplicationsCBC} and stochastic searches in Sec.~\ref{sec:ImplicationsStochastic}. Finally, we summarize and conclude our work in Sec.~\ref{sec:Conclusion}. Throughout this work we adopt the Planck 2015 cosmological parameters with the Hubble constant being $67.9 \, \mathrm{km} \,\mathrm{s}^{-1}\, \mathrm{Mpc}^{-1}$~\cite{Planck:2015fie}.

\section{\label{sec:Networks} Detector Networks}

We consider three detector-network configurations, denoted \emph{Stochastic-optimized}, \emph{CBC-optimized}, and \emph{Balanced}. The detectors sites and sensitivities do not change across all three configurations, which differ only in the orientations of the interferometer arms. 
This allows us to isolate the effect of detector orientation on the performance of the network for different gravitational-wave science cases. For each configuration, we consider both the full network, comprising the two ET detectors and the CE detector, and the corresponding ET-only network obtained by excluding CE. Below, we first describe the detector sites and assumed instrument properties and then outline the procedure used to select the arm orientations.

\subsection{Detector sites and sensitivities}

We consider two ET interferometers located at the candidate sites in Sardinia, Italy, and the Lusatia region of Saxony, Germany, together with a CE interferometer at the fictitious site off the Atlantic coast of Texas considered in Ref.~\cite{Gupta:2023}. 
The ET detectors are modeled as 15-km L-shaped interferometers and adopt the HFLF (low-frequency cryogenic and high-frequency room-temperature xylophone) cryogenic sensitivity curve of Ref.~\cite{Branchesi:2023}. The CE detector is modeled as a 40-km L-shaped interferometer using the sensitivity curve for a 40-km surface detector from Ref.~\cite{Evans:2021}. 
The detector locations and their main properties, including geographical coordinates, arm lengths, and arm orientations, are summarized in Table~\ref{tab:detectors}. The choice of arm orientations is discussed in the following subsection.

\begin{table*}[!ht]
\begin{tabular}{ |l|c|c|c|c|c|c|c| }
 \hline
 Site & Prefix & Arm length & Latitude & Longitude & Elevation & x-arm azimuth & y-arm azimuth\\
 \hline
 \multilinecellleft[c]{Sardinia,\\ Sos Enattos} & \multilinecell[c]{s1\\s2\\s3} & 15 km & $40.472788^\circ$ & $9.448977^\circ$ & 300 m & \multilinecell[c]{$164.814^\circ$ \\  $98.139^\circ$ \\ $174.732^\circ$ } & \multilinecell[c]{$74.814^\circ$ \\ $8.139^\circ$ \\ $84.732^\circ$} \\ 
 \hline
 \multilinecellleft[c]{Saxony,\\ Lusatia} & \multilinecell[c]{l1\\l2\\l3} & 15 km & $51.346692^\circ$ & $14.297119^\circ$ & -100 m & \multilinecell[c]{$168.437^\circ$ \\ $147.714^\circ$ \\ $152.559^\circ$} &  \multilinecell[c]{$78.437^\circ$ \\ $57.714^\circ$ \\ $62.559^\circ$ }\\ 
 \hline
 \multilinecellleft[c]{Atlantic \\ Ocean} & \multilinecell[c]{c1\\c2\\c3} & 40 km & $29.00^\circ$ & $-94.0^\circ$ & 0 m & \multilinecell[c]{ $177.908^\circ$ \\ $90.130^\circ$ \\ $178.416^\circ$} & \multilinecell[c]{$87.908^\circ$ \\ $0.130^\circ$ \\ $88.416^\circ$} \\ 
 \hline
\end{tabular}
\caption{Locations and properties of the detectors considered in this work. Listed are the geographical coordinates, elevation, arm length, and arm orientations of each detector. Arm azimuths are measured clockwise from local North and are given modulo $90^\circ$.} \label{tab:detectors}
\end{table*}

\subsection{Choice of arm orientation}

We choose the interferometer arm orientations following the approach of Ref.~\cite{Ebersold:2024}, considering two science-driven optimization criteria. 
For SGWB searches, we choose the arm orientations that minimize $\Omega_{\mathrm{min}}^{\mathrm{PI}}$, defined as the minimum of the network power-law integrated (PI) sensitivity curve~\cite{Thrane:2013oya}. We refer to the resulting configuration as the \emph{Stochastic-optimized} network.
For CBC observations, we instead choose the arm orientations that maximize the sky-averaged network alignment factor, $\bar{\alpha}$, as defined in Ref.~\cite{Ebersold:2024}. This quantity characterizes the network's ability to distinguish between the plus and cross polarizations of a gravitational-wave signal~\cite{Klimenko:2005xv,Sutton:2009gi}. We refer to the resulting configuration as the \emph{CBC-optimized} network.

\begin{figure}[!htbp]
    \centering
    \includegraphics[width=0.48\textwidth]{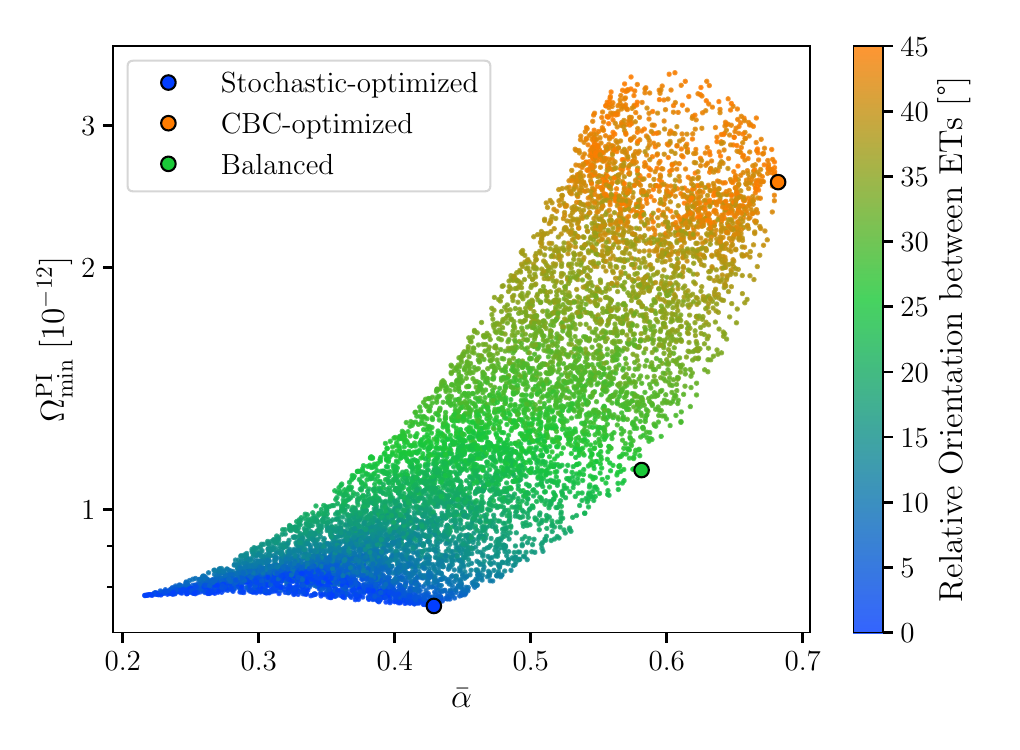}
    \caption{Sky-averaged network alignment factor $\bar \alpha$ versus the minimum of the PI sensitivity curve, $\Omega_\mathrm{min}^\mathrm{PI}$, for random orientations of the two ET and the CE detectors. The color indicates the relative orientation of the two ET interferometers. The Stochastic-optimized, CBC-optimized, and Balanced configurations selected for this work are highlighted.}
    \label{fig:alphaomega}
\end{figure}

To illustrate the trade-off between these two optimization criteria, we evaluate $\Omega_{\mathrm{min}}^{\mathrm{PI}}$ and $\bar{\alpha}$ for a large number of randomly sampled detector orientations. The detector locations, sensitivities and opening angles ($90^\circ$) are kept fixed, while the arm azimuths of all three interferometers are varied independently. The resulting distribution is shown in Fig.~\ref{fig:alphaomega}, where each point represents a network geometry and the color indicates the relative orientation of the two ET interferometers.
The relative ET orientation strongly influences both metrics: approximately aligned ET detectors generally yield better stochastic sensitivity, whereas increasing their relative misalignment leads to a larger network alignment factor. Optimizing the network for one metric therefore generally comes at the expense of the other.
In addition to the two optimized configurations, we select a \emph{Balanced} network representing a compromise between the two science goals. Favorable compromise configurations populate the lower-right region of Fig.~\ref{fig:alphaomega}, corresponding to simultaneously small $\Omega_{\mathrm{min}}^{\mathrm{PI}}$ and large $\bar{\alpha}$. From this region, we select one representative configuration, as indicated in the figure. 
The arm orientations of all interferometers in the three configurations are listed in Table~\ref{tab:detectors}, with azimuths measured clockwise from local North. 

\begin{figure}[!htbp]
    \centering
    \includegraphics[width=0.48\textwidth]{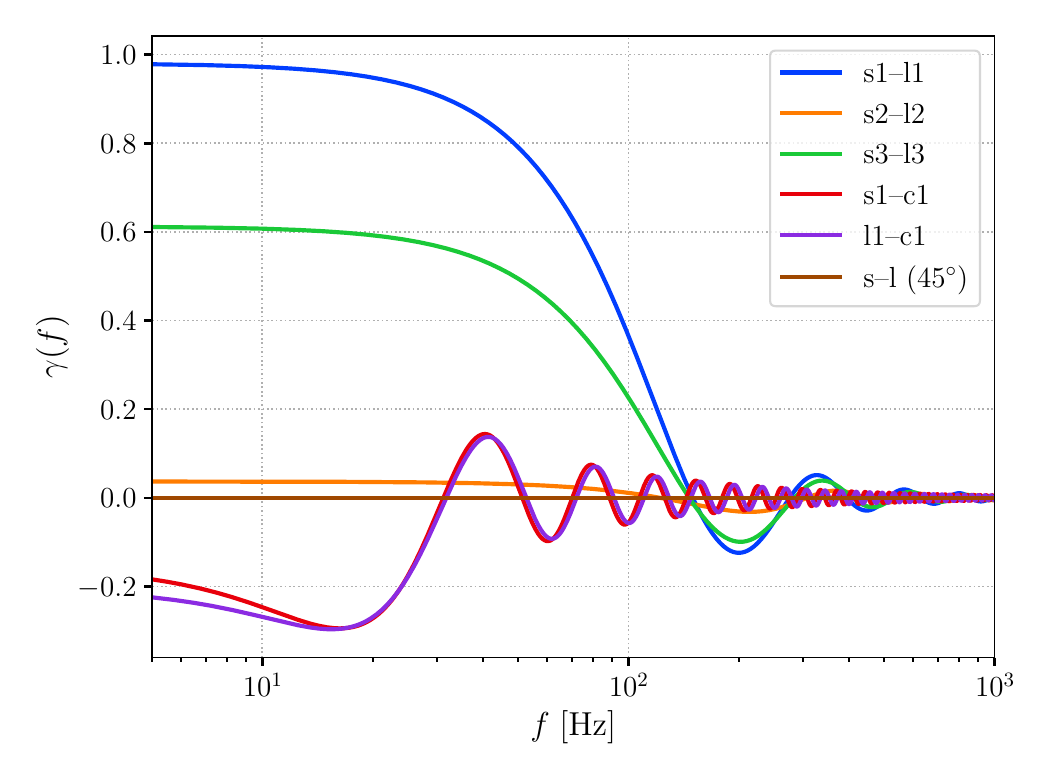}
    \caption{Normalized ORFs for selected detector pairs. For the Sardinia--Saxony baseline, we show the ORFs for the Stochastic-optimized, CBC-optimized, and Balanced configurations, together with a configuration with a relative orientation of exactly $45^\circ$, for which the ORF vanishes completely. 
    For comparison, the CE--Sardinia and CE--Saxony baselines are shown for one network configuration, illustrating the more rapid oscillation associated with their larger detector separationas.}
    \label{fig:orfs}
\end{figure}

The strong dependence of $\Omega_{\mathrm{min}}^{\mathrm{PI}}$ on detector orientation originates from the corresponding overlap reduction functions (ORFs)~\cite{PhysRevD.46.5250,Flanagan:1993ix,PhysRevD.55.448,Christensen:2026yyl}, shown in Fig.~\ref{fig:orfs}. The ORFs characterize the frequency-dependent geometrical sensitivity of each detector pair to an isotropic stochastic background and depend on both the separation and relative orientation of the detectors. 
In the low-frequency limit, the ORF amplitude is governed primarily by the relative detector orientation. Larger detector separations lead to a faster decay of the amplitude and more rapid oscillations with frequency, reflecting the increasing phase differences between the GW signals arriving at the detectors. In particular, aligned detectors exhibit a large low-frequency amplitude, whereas unfavorable relative orientations can strongly suppress it. The different orientation choices can therefore substantially alter the SGWB sensitivity of the network even though the detector locations and noise sensitivities remain unchanged.
The corresponding values of the stochastic and CBC optimization metrics, for both the full network and the ET-only configuration obtained by excluding CE, are summarized in Table~\ref{tab:networks}. These three configurations form the reference networks used throughout the remainder of this work.

\begin{table*}[!ht]
    \centering
    \begin{tabular}{|l|c|c|c|c|c|c|c|}
        \hline
         & \multicolumn{3}{|c|}{Full Network} & \multicolumn{4}{|c|}{ET-only} \\
        \hline
         Configuration & Detectors & $\Omega_{\mathrm{min}}^{\mathrm{PI}}\; [10^{-12}]$ & $\bar \alpha$ & Detectors & $\Omega_{\mathrm{min}}^{\mathrm{PI}}\; [10^{-12}]$ & $\bar \alpha$ & Relative ET angle \\
        \hline
        Stochastic-optimized & s1 l1 c1 & $\omegaStoch$ & $\alphaStoch$ & s1 l1 & $\omegaStochET$ & $\alphaStochET$ & $\angleStochET^\circ$ \\
        CBC-optimized & s2 l2 c2 & $\omegaCBC$ & $\alphaCBC$ & s2 l2 & $\omegaCBCET$ & $\alphaCBCET$ & $\angleCBCET^\circ$ \\
        Balanced & s3 l3 c3 & $\omegaBalance$ & $\alphaBalance$ & s3 l3 & $\omegaBalanceET$ & $\alphaBalanceET$ & $\angleBalanceET^\circ$ \\
    \hline
    \end{tabular}
    \caption{Properties of the three detector-network configurations considered in this work. For each configuration, we list the constituent detectors and the corresponding stochastic and CBC optimization metrics, $\Omega_{\mathrm{min}}^{\mathrm{PI}}$ and $\bar{\alpha}$, for both the full network and the ET-only network. Additionally, the relative orientation of the two ET interferometers is also given.}
    \label{tab:networks}
\end{table*}

\section{\label{sec:ImplicationsCBC} Implications for compact binary searches}

In this section we assess the performance of the six detector networks for CBC searches. We consider three representative source classes: a $1.4 + 1.4 \, M_\odot$ BNS system, and two BBH systems with component masses of $10 + 10 \, M_\odot$ and $30 + 30 \, M_\odot$, respectively. 
We compare the networks in terms of their detection efficiency, sky-localization and luminosity-distance estimation.
Following Ref.~\cite{Ebersold:2024}, we present efficiency curves as a function of redshift, allowing the different orientation configurations to be compared independently of assumptions about the underlying astrophysical source population.

\subsection{Detection efficiency}

For the BNS system, we assume component masses of $1.4\,M_\odot$ and zero spins. The tidal deformabilities are fixed to $\Lambda_1 = \Lambda_2 = 257$, corresponding to the APR4 equation of state~\cite{Akmal:1998}. We model the gravitational-wave signal using \texttt{IMRPhenomXAS\_NRTidalv3}~\cite{Pratten:2020,Abac:2023}, which describes the dominant mode during the inspiral and tapers the signal amplitude to zero above the estimated merger frequency, without including a post-merger signal.
For the BBH systems, we likewise assume non-spinning components and use the phenomenological inspiral-merger-ringdown waveform model \texttt{IMRPhenomXAS}~\cite{Pratten:2020}.

We evaluate the signals over the frequency range $10$--$2048\, \mathrm{Hz}$ for BNSs and $8$--$2048\, \mathrm{Hz}$ for BBHs. For the BNS systems, we restrict the analysis to frequencies above $10\, \mathrm{Hz}$ to keep the signal duration sufficiently short that the time dependence of the detector antenna-pattern functions can be neglected. At lower frequencies, the longer signal would require accounting for the Earth's rotation. This resulting modulation can provide additional localization and polarization information and is therefore expected to improve the performance of all networks in this regard.
Sources are distributed isotropically over the sky, with the cosine of the inclination angle drawn uniformly.

\begin{figure*}[!ht]
    \centering
    \includegraphics[width=0.98\textwidth]{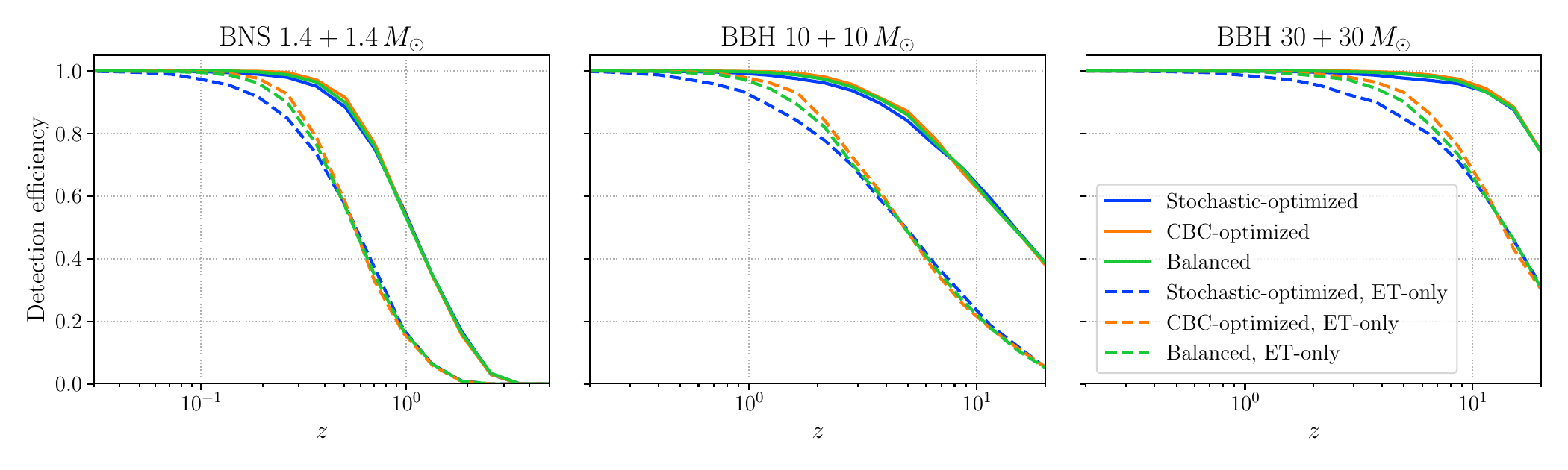}
    \caption{Detection efficiency as a function of redshift for the six detector networks and the three reference CBC systems considered in this study: a $1.4 + 1.4 \, M_\odot$ BNS and $10 + 10 \, M_\odot$ and $30 + 30 \, M_\odot$ BBHs. A signal is considered detected when the optimal network SNR exceeds $\rho_\mathrm{net} = 12 $.}
    \label{fig:detefficiency}
\end{figure*}

We consider a signal to be detected when its optimal network signal-to-noise ratio (SNR) satisfies $\rho_\mathrm{net} > 12$~\cite{KAGRA:2013rdx}. We emphasize that this constitutes an idealized detection criterion based solely on the optimal SNR, rather than a full search pipeline accounting for, e.g., detection-statistic thresholds and non-Gaussian noise transients.
Fig.~\ref{fig:detefficiency} shows the detection efficiency of the six networks for the three reference systems as a function of redshift, restricted to $z \leq 20$.
The dominant difference is between the ET-only and full networks, with the latter reaching substantially larger redshifts. In contrast, changes in detector orientation have only a modest impact. The largest differences occur near the shoulder of the efficiency curves, where the stochastic-optimized geometries begin to lose detection efficiency at somewhat lower redshifts than their CBC-optimized counterparts.

\subsection{Sky localization}

Beyond the detection of CBCs, accurately localizing their position on the sky is crucial for exploiting their astrophysical and cosmological information. For BNS mergers, precise localization enables electromagnetic follow-up observations and the identification of host galaxies~\cite{LIGOScientific:2017ync}. For BBHs, which are generally not expected to produce an electromagnetic counterpart, accurate localization facilitates statistical associations with host galaxies and environments, providing insight into their formation channels and population properties~\cite{Fan:2014kka,Mapelli:2021}. Accurate CBC source localization can furthermore aid cosmological measurements with gravitational-wave standard sirens, including determinations of the Hubble constant~\cite{LIGOScientific:2017adf,DES:2019ccw,LIGOScientific:2026uyd}.

\begin{figure*}[!ht]
    \centering
    \includegraphics[width=0.98\textwidth]{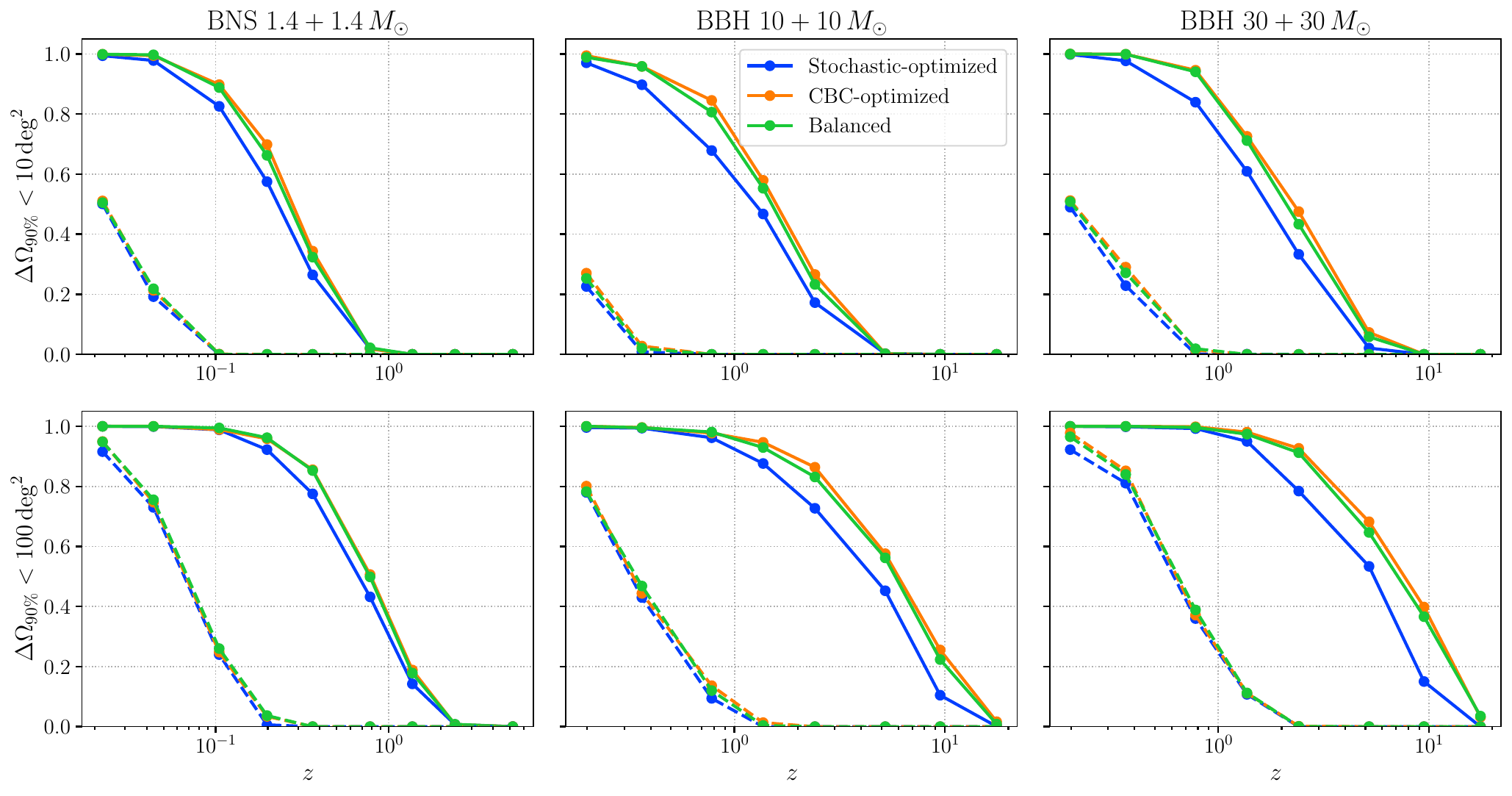}
    \caption{Localization efficiency as a function of redshift for BNS systems (left), $10 + 10 \, M_\odot$ BBHs (middle) and $30 + 30 \, M_\odot$ BBHs (right). The top (bottom) row shows the fraction of events with a reconstructed $90\%$ credible sky area smaller than $10\,\mathrm{deg}^2$ ($100\,\mathrm{deg}^2$). Solid curves denote the full ET + CE networks, while dashed curves correspond to the ET-only configurations. More than 1000 randomly orientated sources are injected at each redshift. The addition of a third detector leads to a substantial improvement in localization efficiency across all source classes and redshifts, whereas the dependence on the detector orientation is comparatively weak.}
    \label{fig:localization}
\end{figure*}

We therefore investigate the sky-localization capabilities of the different detector networks for our reference systems.
Following Ref.~\cite{Ebersold:2024}, we estimate the source parameters using the single template model implemented in \texttt{PyCBC\_inference}~\cite{Biwer:2018}. For each simulated signal, a single reference waveform is generated with the intrinsic parameters fixed to their injected values, while sampling only over the extrinsic source parameters. This approach substantially reduces the computational cost of parameter estimation as the relatively expensive waveform generation is performed only once and the dimensionality of the parameter space is reduced. 
Although this approximation does not capture the full diversity of the parameters, 
%correlations with intrinsic source parameters, 
it is sufficient for assessing the relative localization performance of the different detector networks.
From the resulting posterior distributions in right ascension and declination, we estimate the sky area enclosing 90\% of the posterior probability~\cite{Singer:2015ema}. Analogously to the detection efficiency, we then compute the fraction of events localized within $10\,\mathrm{deg}^2$ and $100\,\mathrm{deg}^2$ as a function of redshift. The resulting localization efficiencies are shown in Fig.~\ref{fig:localization}.

The localization performance shows a pronounced difference between the ET-only and full three-detector networks, whereas the differences between the three orientation configurations are comparatively small. This demonstrates that the presence of a third detector is the dominant factor determining localization performance. The primary localization mechanism is triangulation: the large geographical separation between ET and CE introduces measurable arrival-time differences that strongly constrain the source position on the sky. Differences between the orientation configurations enter only at second order through their sensitivity to the gravitational-wave polarization content~\cite{Wen:2010cr,Fairhurst:2010is,Fairhurst:2017mvj}.
For the full networks, essentially all nearby ($z<0.1$) BNS mergers can be localized within $100\,\mathrm{deg}^2$. In contrast, this fraction is only about $25\%$ for the ET-only configurations, largely independent of the detector orientation. 
A similar behavior is observed for BBHs, with the full networks consistently outperforming the corresponding ET-only configurations across the entire redshift range.

To illustrate the impact of the network configuration on the localization of an individual event, we consider a representative $10 + 10 \, M_\odot$ BBH at a luminosity distance of $10\,\mathrm{Gpc}$, corresponding to a redshift of $z=1.37$.
The choice of sky position requires some care, since the localization performance depends on the source position relative to the detector antenna patterns and could therefore favor a particular orientation configuration.
We therefore select an event whose reconstructed $90\%$ credible sky area, $\Delta \Omega_{90\%}$, lies close to the median value at this redshift for all network configurations. The event is thus representative of the typical localization performance of the considered networks.
The network SNRs are also similar among the different orientation configurations, with $\rho_\mathrm{net} \simeq 50$ for the ET-only networks and $\rho_\mathrm{net} \simeq 75$ for the corresponding full networks.

\begin{figure*}[!ht]
    \centering
    \includegraphics[width=0.93\textwidth]{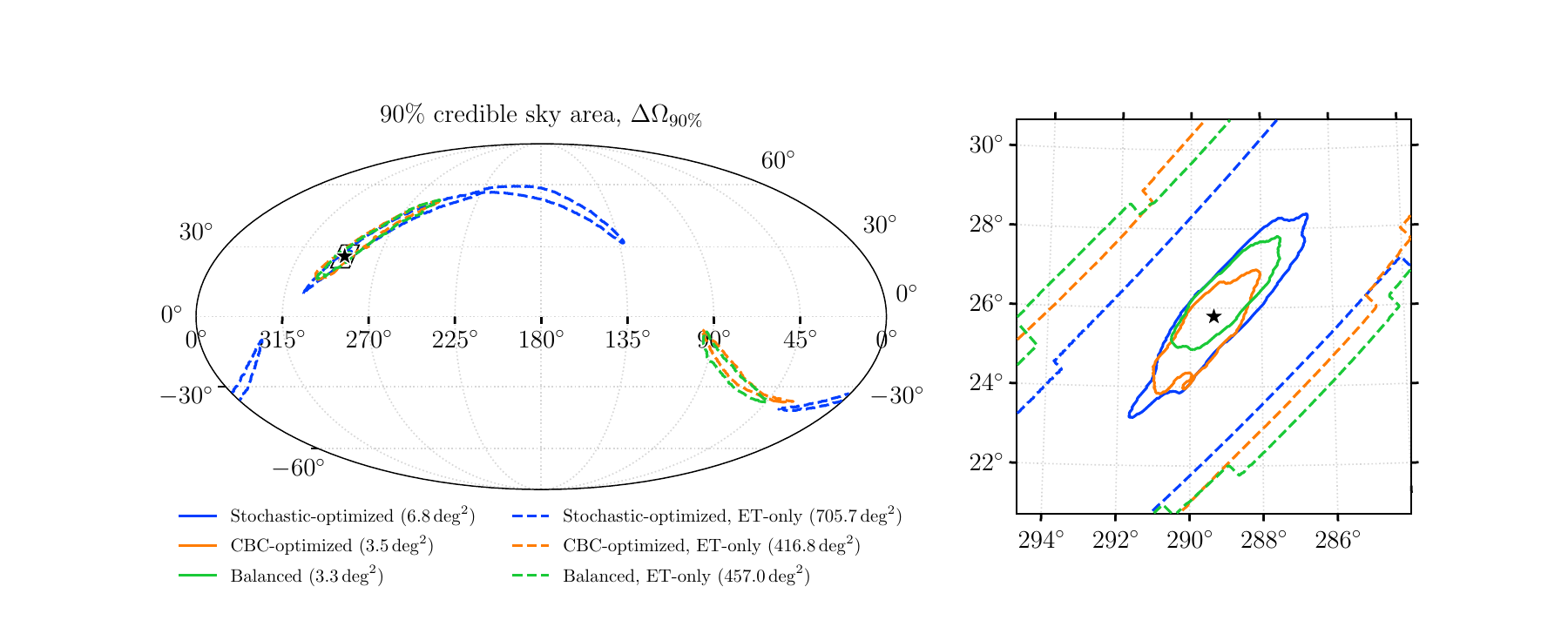}
    \caption{Sky localization for a representative $10+10\,M_\odot$ BBH injection at a luminosity distance of $10\,\mathrm{Gpc}$ ($z=1.37$). The contours enclose the $90\%$ credible sky regions for the Stochastic-optimized, CBC-optimized, and Balanced configurations, for both the ET-only (dashed lines) and full ET+CE (solid lines) networks. 
    The corresponding credible areas $\Delta\Omega_{90\%}$ are given in the legend, and the star marks the injected sky position. The right panel provides a zoomed view around the injected position, highlighting the substantially smaller localization regions obtained when CE is added to the network.}  \label{fig:skymaps}
\end{figure*}

The resulting $90\%$ credible regions are shown in Fig.~\ref{fig:skymaps}. For the ET-only configurations, the reconstructed sky areas are $706 \, \mathrm{deg}^2$, $417 \, \mathrm{deg}^2$, and $457 \, \mathrm{deg}^2$ for the stochastic-optimized, CBC-optimized, and balanced networks, respectively. The corresponding full networks reduce these areas to approximately $7 \, \mathrm{deg}^2$, $4 \, \mathrm{deg}^2$, and $3 \, \mathrm{deg}^2$, representing an improvement of roughly two orders of magnitude.
The extended banana-shaped credible regions obtained with the ET-only networks collapse into compact sky patches once the third detector (i.e. the CE detector) is added. Although the CBC-optimized and balanced configurations achieve slightly smaller localization areas than the stochastic-optimized configuration, these differences are modest compared to the dramatic improvement provided by triangulation in the three-detector networks.
We note, however, that the exact sky areas obtained for a single event depend sensitively on the particular noise realization and signal duration (and hence also Earth rotation~\cite{Santoliquido:2026bej}), and our examples should therefore be regarded as illustrative rather than robust measures of average network performance. The localization efficiencies shown in Fig.~\ref{fig:localization} provide a more representative characterization of the overall network capabilities.

\subsection{Distance estimation}

Gravitational-wave observations provide a direct measurement of the luminosity distance to CBC sources. Combined with redshift information, these measurements enable studies of the cosmic expansion history.
For events with an identified electromagnetic counterpart or host galaxy, this forms the basis of the bright-standard-siren method~\cite{LIGOScientific:2017adf}, while statistical associations with galaxy catalogs enable analogous dark-siren measurements for sources without a unique host identification~\cite{LIGOScientific:2026uyd}. Improving the precision of gravitational-wave distance measurements therefore directly enhances the constraining power of such analyses.
A fundamental limitation of gravitational-wave distance measurements is the degeneracy between luminosity distance and binary inclination~\cite{Cutler:1994ys}. The observed gravitational-wave amplitude scales inversely with luminosity distance, while the binary inclination determines the relative amplitudes of the two gravitational-wave polarizations. Therefore, a detector network with an improved ability to resolve the two polarizations can reduce this degeneracy and thereby improve the luminosity distance measurement.

\begin{figure*}[!ht]
    \centering
    \includegraphics[width=0.98\textwidth]{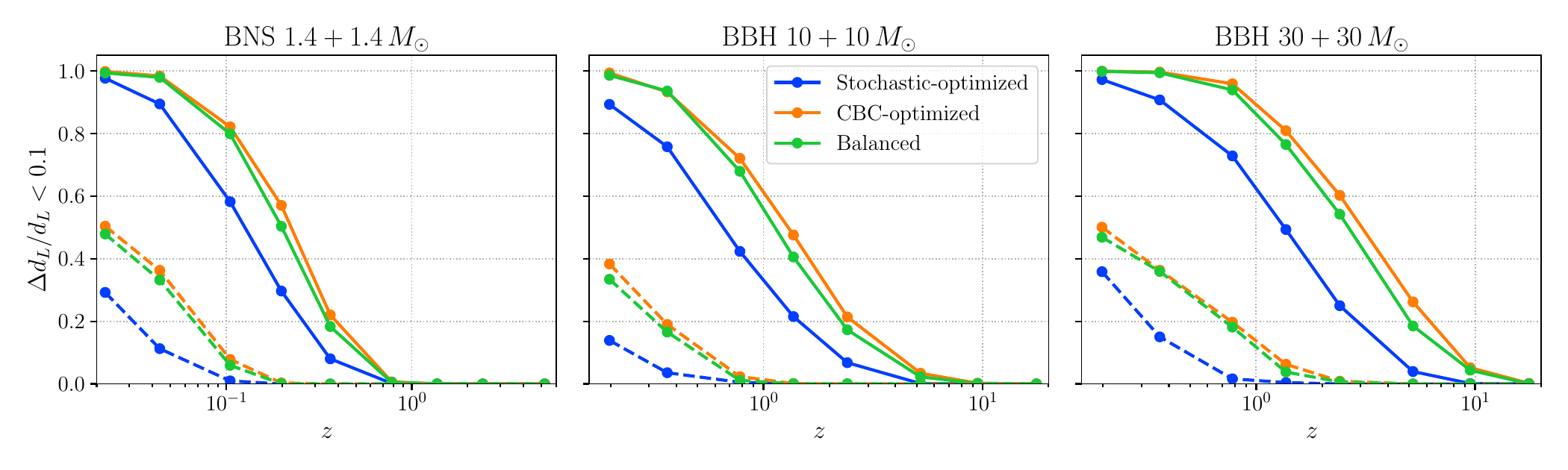}
    \caption{Luminosity-distance estimation performance for the six detector networks. The curves show the fraction of detected events with a fractional luminosity-distance uncertainty below $10\%$ as a function of redshift. While the addition of CE substantially improves the distance measurement for all source classes, detector orientation also plays an important role, with the CBC-optimized and balanced configurations yielding tighter distance constraints than the stochastic-optimized configurations.}
    \label{fig:distance}
\end{figure*}

We estimate the luminosity distance using the same PyCBC parameter-estimation analysis employed for the sky-localization study. For each detected injection, we quantify the (1 $\sigma$) fractional distance uncertainty as $\Delta d_L / d_L =(d_L^{84\%} - d_L^{16\%})/ (2 d_L)$, where $d_L^{16\%}$ and $d_L^{84\%}$ denote the 16th and 84th percentiles of the marginalized luminosity-distance posterior, respectively, and $d_L$ is the injected luminosity distance.
Analogously to the localization analysis, we then evaluate the fraction of detected events for which the luminosity distance is recovered with a fractional uncertainty below $10\%$ as a function of redshift.
Fig.~\ref{fig:distance} shows this fraction for all reference source systems and network configurations. For all sources, the addition of CE leads to a substantial improvement in distance estimation compared with the corresponding ET-only networks.
In contrast to the sky-localization results, however, the distance-measurement performance exhibits a much stronger dependence on the detector orientation. This reflects the fact that luminosity-distance estimation benefits from disentangling the two gravitational-wave polarizations and is therefore more sensitive to the polarization response of the network.
As expected from its optimization for polarization sensitivity, the CBC-optimized configuration consistently provides the best distance measurements, while the Stochastic-optimized configuration performs less favorably, for both the ET-only and full networks. Interestingly, the Balanced configuration retains most of the distance-estimation performance of the CBC-optimized network.
This indicates that a network geometry providing a compromise between stochastic background and CBC optimization can retain much of the improvement in distance estimation associated with enhanced polarization sensitivity.

\begin{figure}[!htbp]
    \centering
    \includegraphics[width=0.48\textwidth]{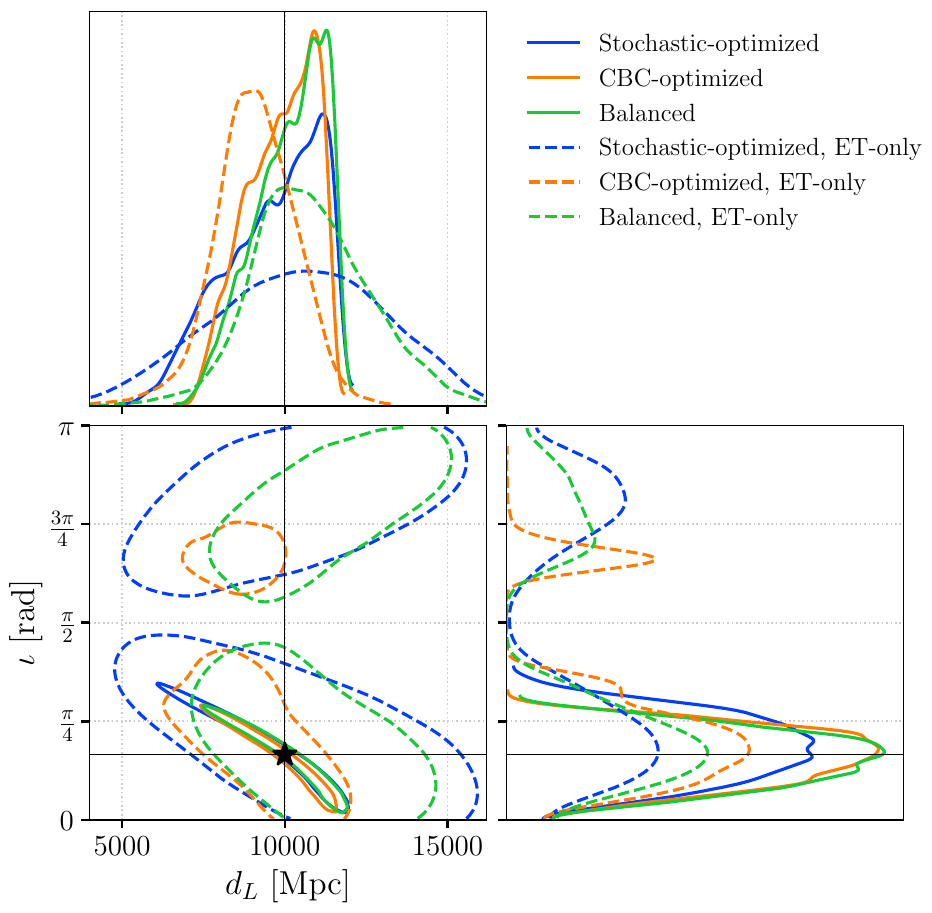}
    \caption{Joint and marginalized posterior distribution of luminosity distance, $d_L$, and inclination angle, $\iota$, for the representative BBH event also used for Fig.~\ref{fig:skymaps}. Solid contours denote the full ET + CE detector network, while dashed contours correspond to the two ET detectors alone for each orientation configuration. Contours enclose $90\%$ of the posterior probability. The injected parameters are indicated by the black star.  }
    \label{fig:distanceInclination}
\end{figure}

To illustrate the origin of these trends, we consider the posterior distributions obtained for the same representative event used in the sky-localization study. The injected inclination is $\iota = 0.52$. In Fig.~\ref{fig:distanceInclination} we show the joint and marginalized posterior distributions of the inclination angle and luminosity distance. As in the localization example, the figure should be regarded as an illustration of the typical posterior structure, since the detailed morphology depends on the specific source parameters and noise realization. 
The figure illustrates the well-known degeneracy between the luminosity distance and the inclination angle, which arises because different combinations of these parameters can produce similar changes to the gravitational-wave amplitudes. The three network configurations yield broadly consistent posterior maxima close to the injected values, indicated by the black star. However, significant differences are visible in the extent of the posterior distributions. 

The full three-detector networks constrain both the luminosity distance and inclination angle substantially better than the corresponding ET-only configurations, highlighting the importance of the additional baselines and polarization information provided by the CE detector. 
In addition to reducing the overall posterior width, the inclusion of the CE detector largely resolves the inclination-angle degeneracy present in the ET-only configurations. While the ET-only networks exhibit two posterior modes corresponding to the $\iota$ and $\pi-\iota$ solutions, the full networks strongly favor the injected low-inclination configuration.
Among the considered layouts, the CBC-optimized and balanced configurations produce the tightest constraints, whereas the stochastic-optimized network exhibits broader posteriors, particularly in luminosity distance.
These differences are also evident in the marginalized one-dimensional posteriors, demonstrating how the detector geometry influences the ability of the network to constrain both the binary inclination and luminosity distance.

\section{\label{sec:ImplicationsStochastic} Implications for stochastic searches}

We assess the implications of the different network geometries for SGWB searches in three steps. First, we compare their broadband sensitivity to stochastic backgrounds. We then investigate their ability to constrain an additional cosmological background in the presence of the astrophysical CBC background, before assessing the ability to distinguish such a component through Bayesian model selection.

\subsection{PI curves}

\begin{figure}[!htbp]
    \centering
    \includegraphics[width=0.48\textwidth]{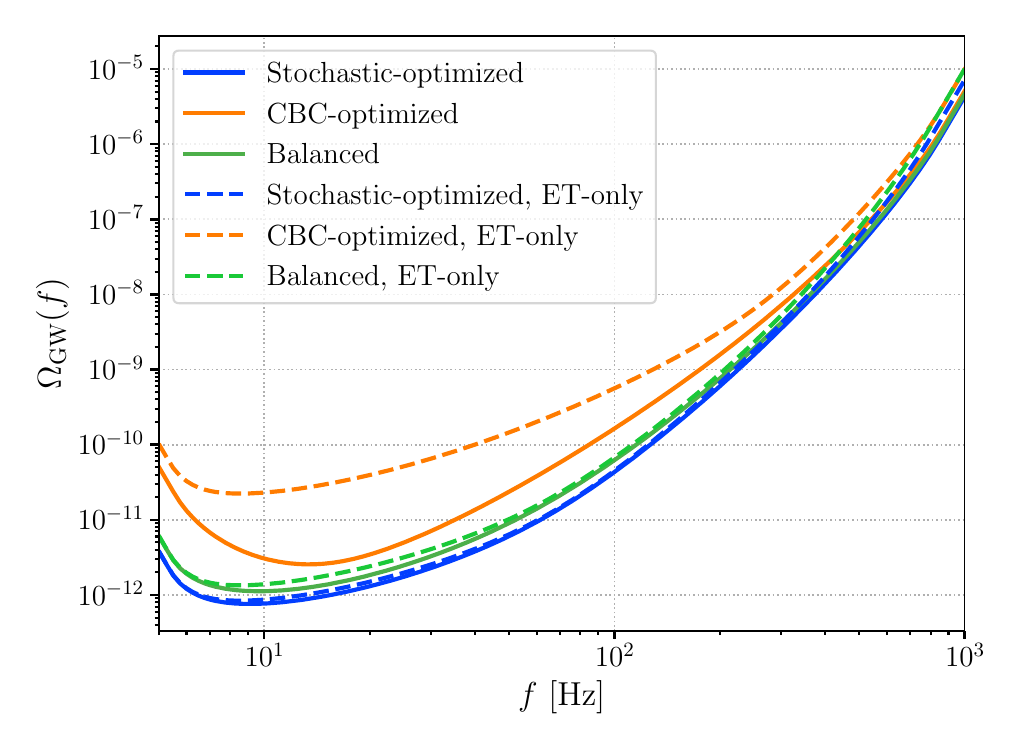}
    \caption{PI curves for all network configurations, assuming one year of coincident observing time and $\mathrm{SNR} = 3$.}
    \label{fig:PIcurves}
\end{figure}

Fig.~\ref{fig:PIcurves} shows the PI sensitivity curves for our network configurations, assuming one year of coincident observation time and ${\rm SNR} = 3$. These curves provide a convenient measure of the broadband SGWB sensitivity of a detector network to power-law backgrounds~\cite{Thrane:2013oya}.
As expected, the Stochastic-optimized configuration provides the best broadband sensitivity. Remarkably, the stochastic-optimized ET-only network already achieves a sensitivity close to that of the full network.
In contrast, adding CE to the nearly misaligned CBC-optimized ET configuration recovers some of the lost SGWB sensitivity. Nevertheless, the resulting full network remains less sensitive than the ET-only Balanced configuration, with the difference being particularly pronounced at low frequencies. 
For the ET only Balanced configuration that we have studied, with a relative alignment of $\angleBalanceET^\circ$, the sensitivity to the energy density of the SGWB is 62.5\% of the ideal configuration, 0$^\circ$. For a relative alignment of 30$^\circ$~\cite{CERN:EDMS-3529996} the sensitivity drops to 50\%, while for 22.5$^\circ$ it is 70.7\%~\cite{Christensen:2026yyl}. 

The minimum of the PI curve, $\Omega_\mathrm{min}^\mathrm{PI}$, for the full CBC-optimized network is more than a factor of three higher than for the stochastic-optimized network. Since the stochastic SNR scales as $\sqrt{T{\rm obs}}$, recovering a factor of more than three in amplitude sensitivity would require more than an order of magnitude increase in observing time, from the one year assumed here to more than ten years. 
Moreover, the difference between the configurations is not limited to the minima of the PI curves: the CBC-optimized configuration has poorer broadband sensitivity, particularly at low frequencies. Matching the stochastic-optimized configuration for a background extending over a broad frequency range could therefore require an even larger increase in observing time.

While PI curves provide an intuitive comparison of the stochastic sensitivity of different detector networks, they do not directly describe the ability of a network to separate multiple contributions to the SGWB. In particular, a search for a cosmological background must distinguish such a component from the astrophysical background produced by CBCs. We therefore next investigate this problem through joint parameter estimation of the astrophysical and cosmological SGWB components.

\subsection{Upper limits on a cosmological background}

We follow the approach of Ref.~\cite{Martinovic:2020} and perform a simultaneous estimation of astrophysical and cosmological contributions to the SGWB.
The background is modeled as the sum of a CBC component with the characteristic inspiral spectrum of a CBC background, $\Omega_{\rm CBC}(f) \propto f^{2/3}$, and a flat cosmological component $\Omega_\mathrm{Cosmo}$,
\begin{align}
    \Omega_\mathrm{GW}(f) = \Omega_\mathrm{CBC}^\mathrm{ref} \left(\frac{f}{f_\mathrm{ref}}\right)^{2/3} + \Omega_\mathrm{Cosmo}\,,
    \label{eq:cbc_cosmo_model}
\end{align}
where $\Omega_\mathrm{CBC}^\mathrm{ref}$ denotes the amplitude of the CBC component at the reference frequency, for which we choose  $f_\mathrm{ref} = 10 \, \mathrm{Hz}$ throughout. The analysis is restricted to the frequency range $5$--$100 \,\mathrm{Hz}$, where the CBC background is well approximated by a power law~\cite{Regimbau:2011rp,Ebersold:2025izh}.
A flat spectrum provides a simple phenomenological model for a cosmological contribution and is, for example, predicted for cosmic strings in the relevant frequency range~\cite{Siemens:2006yp,Gouttenoire:2019kij}.

\begin{figure}[!htbp]
    \centering
    \includegraphics[width=0.48\textwidth]{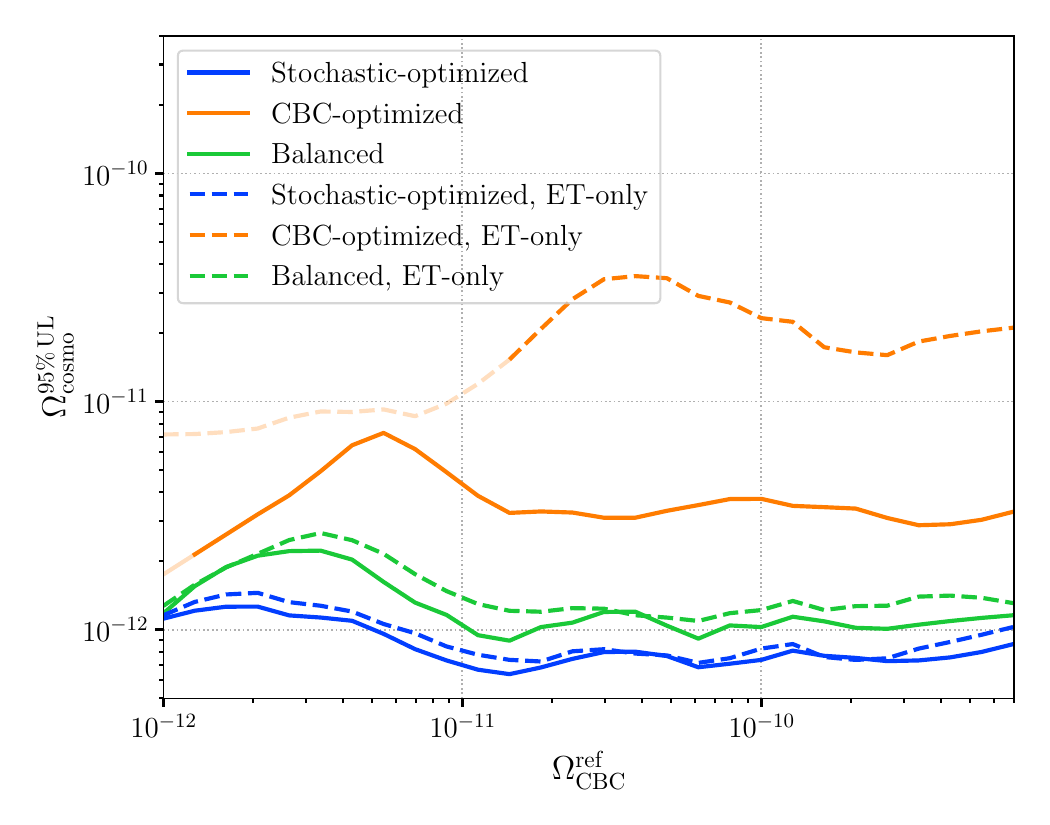}
    \caption{$95\%$ credible upper limits on a flat cosmological SGWB component, as a function of the injected CBC background amplitude. Curves are shown with reduced opacity where the stochastic-signal model is not favored over noise.}
    \label{fig:CosmoUpperLimits}
\end{figure}

To determine how strongly an additional flat component can be constrained in the presence of a CBC background, we inject a range of values for $\Omega_\mathrm{CBC}^\mathrm{ref}$ while setting $\Omega_\mathrm{Cosmo} = 0$. The simulated data are analyzed using the full two-component model of Eq.~\ref{eq:cbc_cosmo_model} with a 
stochastic cross-correlation likelihood and log-uniform priors on both amplitudes between $10^{-15}$ and $10^{-8}$. For each network configuration, we determine the $95\%$ credible upper limit on $\Omega_\mathrm{Cosmo}$. 
Fig.~\ref{fig:CosmoUpperLimits} shows the resulting constraints. The stochastic-optimized configuration consistently yields the most stringent upper limits on the flat component, with only a modest difference between the ET-only and the full ET+CE networks. The Balanced configuration performs comparably well, whereas the CBC-optimized configuration yields weaker constraints, particularly in the ET-only case.

Given the LIGO-Virgo CBC observations, it is expected that there is an astrophysical SGWB at the level of 
$\Omega_\mathrm{CBC}(25 ~ \mathrm{Hz}) = 6.3^{+5.0}_{-2.2} \times 10^{-10}$ at the 90\% credible level~\cite{LIGOScientific:2026mjf}. Converting this to the 10 Hz reference frequency we use in this study, $\Omega^\mathrm{ref}_\mathrm{CBC} \approx 3.4 \times 10^{-10}$. It could be possible to reduce $\Omega^\mathrm{ref}_\mathrm{CBC}$ in the search for $\Omega_\mathrm{Cosmo}$ by removing individually detected CBC signals, as described in, for example, Refs.~\cite{Regimbau:2016ike,Sachdev:2020bkk,Zhong:2022ylh}. This is why we reduce the value of 
$\Omega^\mathrm{ref}_\mathrm{CBC}$ in Fig.~\ref{fig:CosmoUpperLimits}.

The behavior of the upper-limit curves can be understood in terms of the CBC background amplitude and the stochastic sensitivity of the networks. In the regime where no stochastic signal is favored over noise, the upper limit is determined primarily by the stochastic sensitivity of the network 
and is therefore approximately constant. 
Note that in Fig.~\ref{fig:CosmoUpperLimits}, this regime is only visible for the CBC-optimized ET-only network, where the limit on $\Omega_{\rm Cosmo}^{95\%}$ is $\sim 10^{-11}$ as $\Omega_{\rm CBC}^{\rm ref}$ increases from $10^{-12}$ to $10^{-11}$. 
As $\Omega_\mathrm{CBC}^\mathrm{ref}$ increases, the CBC background becomes detectable. However, the total stochastic power is constrained more strongly than its decomposition into CBC and flat components. The two components are degenerate, causing the upper limit on $\Omega_{\rm Cosmo}^{95\%}$ to increase roughly in proportion to $\Omega_{\rm CBC}^{\rm ref}$. 
At still larger CBC background amplitudes, it is measured with sufficiently high SNR that its spectral shape can be distinguished from that of a flat component. The degeneracy between the two contributions is consequently broken, and the upper limits decrease before flattening. The asymptotic level reached in this regime reflects the ability of a network to separate the two spectral components.

These results reinforce the picture obtained from the PI curves. Adding CE to an ET configuration that is already favorable for stochastic searches provides only a modest improvement in the constraint on a flat component. In contrast, CE provides a substantial gain when the ET-only geometry has poor stochastic sensitivity. Nevertheless, the resulting constraints remain weaker than those obtained when the ET detectors themselves are oriented more favorably for stochastic searches. Thus, the loss of stochastic sensitivity caused by the relative ET orientation cannot be fully compensated by the addition of CE.

\subsection{Distinguishing a cosmological background}

We finally address the stronger question of whether an additional cosmological component can not only be constrained, but also distinguished from the CBC background. Following Ref.~\cite{Martinovic:2020}, we use Bayesian model selection to compare a CBC-only model, $\mathcal{M}_{\rm CBC}$, with a two-component model containing both a CBC and a flat cosmological background, $\mathcal{M}_{\rm CBC+Cosmo}$.
For each simulated cross-correlation spectrum, we perform parameter estimation under both hypothesis and compute the logarithmic Bayes factor 
\begin{align}
    \ln \mathcal{B}^\mathrm{CBC+Cosmo}_\mathrm{CBC} = \ln Z_\mathrm{CBC+Cosmo} - \ln Z_\mathrm{CBC}\,,
    \label{eq:bayes_factor_cosmo}
\end{align}
where $Z_\mathcal{M}$ denotes the Bayesian evidence for model $\mathcal{M}$. We adopt the same priors as in the upper-limit analysis described above.

\begin{figure*}[!t]
    \centering
    \includegraphics[width=0.98\textwidth]{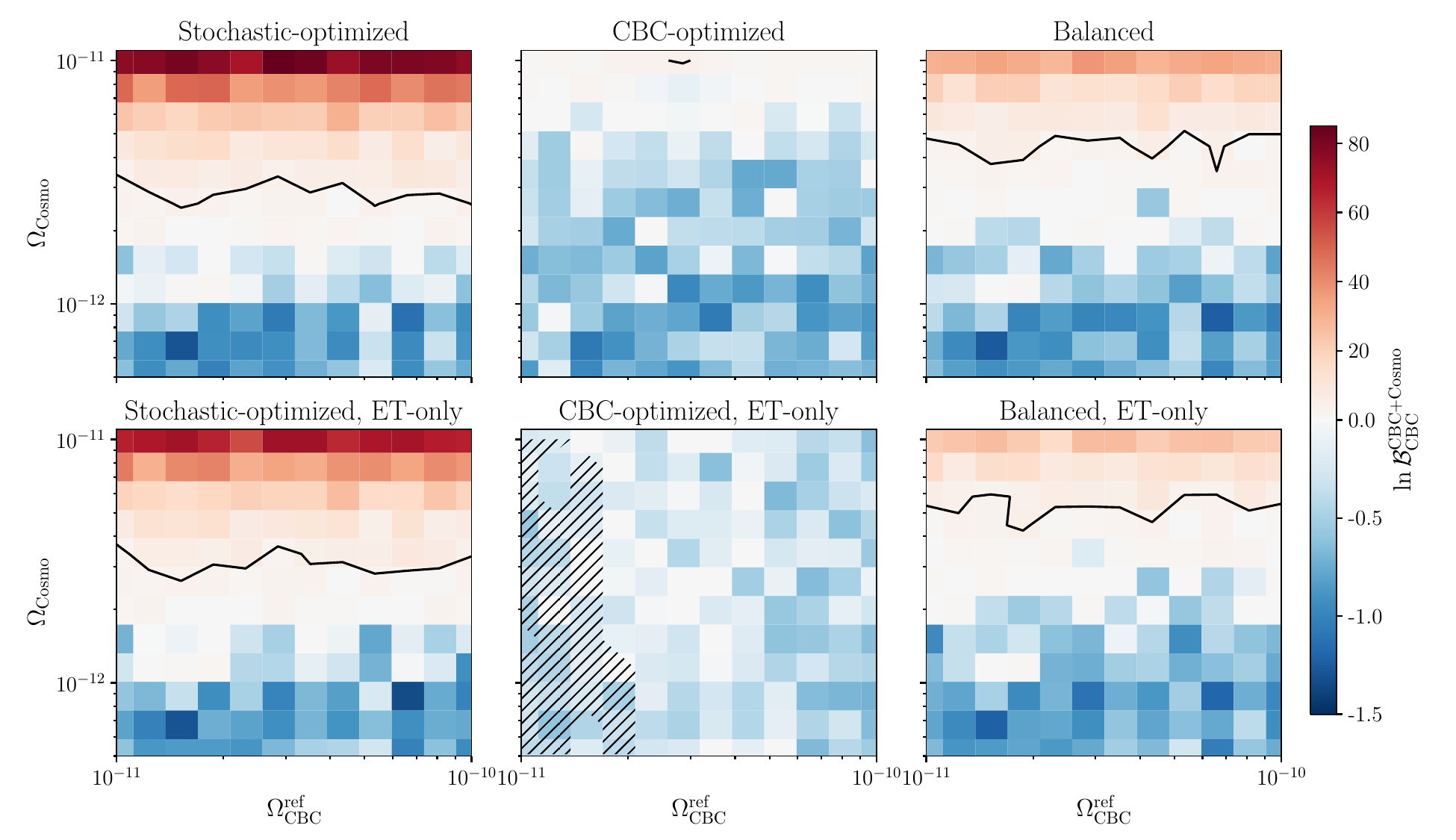}
    \caption{The logarithmic Bayes factor favoring the CBC+Cosmo hypothesis over the CBC-only model is indicated by the color as a function of injected amplitude of $\Omega_{\rm Cosmo}$ and $\Omega_{\rm CBC}^\mathrm{ref}$. The $\ln \mathcal{B}^\mathrm{CBC+Cosmo}_\mathrm{CBC} = 4$ contour is drawn by a black line, which is roughly when we start seeing significant difference between the models. 
    Hatched regions indicate injections for which at least one of the CBC-only or CBC+Cosmo hypotheses is not favored over noise.}
    \label{fig:CBCcosmo}
\end{figure*}

To enable a direct comparison between detector geometries, we use matched noise realizations across all network configurations.
We evaluate the Bayes factor on a grid of injected $\Omega_{\rm CBC}^{\rm ref}$ and $\Omega_{\rm Cosmo}$ amplitudes. The results are shown for each network in Fig.~\ref{fig:CBCcosmo}.
Negative values of $\ln \mathcal{B}_ \mathrm{CBC+Cosmo / CBC}$ indicate a preference for the CBC-only model, whereas positive values favor the presence of an additional flat component. Regions in which neither signal model is favored over noise are indicated by hatching.
The Stochastic-optimized network shows the strongest preference for an additional flat  cosmological component, becoming sensitive to amplitudes above approximately $\Omega_{\rm Cosmo} \simeq 2 \times 10^{-12}$. The transition between the CBC-only and CBC+Cosmo models depends only weakly on the injected CBC amplitude over the range considered here. This is consistent with the upper-limit results of the previous section: once the CBC contribution can be sufficiently well characterized, its presence does not substantially limit the network's ability to identify an additional component with a distinct spectral shape.
The Balanced configuration shows qualitatively similar behavior, albeit requiring larger cosmological amplitudes to favor the two-component model. For the CBC-optimized configuration, considerably larger cosmological amplitudes are required, demonstrating that the loss in stochastic sensitivity translates directly into a reduced ability to identify a subdominant cosmological contribution.

The model-selection analysis provides a more thorough measure of the scientific impact of the detector geometry than the PI curves alone. While PI curves quantify broadband sensitivity to a stochastic signal, the Bayes factors quantify whether the data contain sufficient information to distinguish an additional cosmological component from the CBC background. 
For the simple flat spectrum considered here, we find that the CBC amplitude has only a modest effect on this distinction. This result relies in part on the different spectral shapes of the two components. Cosmological backgrounds with more complex spectra, or with spectral shapes that more closely resemble the CBC background across the sensitive frequency range, are expected to be more difficult to distinguish.

\section{\label{sec:Conclusion} Conclusion}

In this study we have demonstrated the importance of detector orientation for two ET L-shaped detectors. A relative orientation of $45^\circ$ (defined with respect to the great circle connecting the detector sites) effectively eliminates sensitivity to a SGWB, whereas a relative orientation of $0^\circ$ is optimal for stochastic searches. Such a configurations somewhat diminishes in particular the luminosity distance estimation for CBC sources; sky localization remains equally relatively poor regardless of alignment. Positively, the inclusion of a single CE detector located in the US will greatly improve the detectability and localization of CBC sources. This improvement can be achieved while retaining strong SGWB performance, provided that the two ET detectors are appropriately aligned. Notably, we have demonstrated that a balanced detector alignment is possible.

In Sec.~\ref{sec:ImplicationsStochastic}, the PI curves, upper limits, and Bayesian model-selection results consistently demonstrate the importance of the relative detector orientations for stochastic searches. The stochastic-optimized configuration provides the strongest performance throughout, while the Balanced configuration retains a substantial fraction of this sensitivity. In contrast, the CBC-optimized configuration suffers a significant loss in stochastic sensitivity, which limits both the achievable constraints and the ability to distinguish an additional cosmological contribution from the astrophysical background.
Interestingly, the Balanced configuration retains most of the distance-estimation performance of the CBC-optimized network despite being partially optimized for stochastic searches. This result suggests that substantial gains for stochastic searches can be achieved without a comparable loss in CBC parameter-estimation performance.

A decision on the configuration of ET is approaching~\cite{CERN:EDMS-3529996}.
As presented in our work, it is important to consider the presence of a CE detector in the US as part of a global network planning strategy. Since next-generation detectors are designed to operate for $\sim 50 $~years, carefully evaluating detector locations and orientations within a global gravitational-wave detector network is well warranted. 
%\appendix
%\section{}

\begin{acknowledgments}
We thank Marie Anne Bizouard, Federico De Lillo, Joe Romano, Alba Romero, Mairi Sakellariadou, Jishnu Suresh, Nick van Remortel and Anuradha Samajdar for useful conversations and input. 
ME is supported by UZH Postdoc Grant No. [FK-25-109].
We thank the Physik-Institut of the University of Zurich for providing computational
resources.
\end{acknowledgments}

\bibliography{references.bib}

\end{document}

%% file: Macros/parameters.tex
\newcommand{\alphaStoch}{\ensuremath 0.43}
\newcommand{\alphaCBC}{\ensuremath 0.68}
\newcommand{\alphaBalance}{\ensuremath 0.58}
\newcommand{\omegaStoch}{\ensuremath 0.76}
\newcommand{\omegaCBC}{\ensuremath 2.55}
\newcommand{\omegaBalance}{\ensuremath 1.12}
\newcommand{\alphaStochET}{\ensuremath 0.07}
\newcommand{\alphaCBCET}{\ensuremath 0.72}
\newcommand{\alphaBalanceET}{\ensuremath 0.43}
\newcommand{\omegaStochET}{\ensuremath 0.84}
\newcommand{\omegaCBCET}{\ensuremath 22.32}
\newcommand{\omegaBalanceET}{\ensuremath 1.34}
\newcommand{\angleStochET}{\ensuremath 0.12}
\newcommand{\angleCBCET}{\ensuremath 43.92}
\newcommand{\angleBalanceET}{\ensuremath 25.67}